\documentclass{iau}
\usepackage{graphicx}
\def\hd5{HD\,51844}
\def\ds{$\delta$ Sct}
\def\cd{d$^{-1}$}
\def\lg{log $g$}

\def\Teff{T$_{\mathrm{eff}}$}
\def\vsini{$v\sin i$}
\def\kms{km\,s$^{-1}$}

   \title[HD\,51844: a $\delta$ Sct binary]
   {CoRoT target HD\,51844: a $\delta$ Scuti star in a binary system with periastron brightening\thanks{Based on observations made with the Mercator Telescope, operated on the island of La Palma by the Flemish Community, at the Spanish Observatorio del Roque de los Muchachos of the Instituto de Astrof\'isica de Canarias. Based on observations obtained with the HERMES spectrograph, which is supported by the Fund for Scientific Research of Flanders (FWO), Belgium , the Research Council of K.U.Leuven, Belgium, the Fonds National Recherches Scientific (FNRS), Belgium, the Royal Observatory of Belgium, the Observatoire de Gen\`eve, Switzerland and the Th{\"u}ringer Landessternwarte Tautenburg, Germany.
    Based on CoRoT space-based photometric data; the CoRoT space mission was developed and operated by the French space agency CNES, with the participation of ESA’s RSSD and Science Programmes, Austria, Belgium, Brazil, Germany, and Spain. Based on observations collected at La Silla Observatory, ESO (Chile) with the HARPS spectrograph at the 3.6-m telescope, under programme LP185.D-0056.  }}
{}
   \author[Markus Hareter \& Margit Papar{\'o}]   %% give here short author list %%
{Markus Hareter$^1$
 \and Margit Papar{\'o}$^1$}
\affiliation{$^1$Konkoly Observatory MTA CSFK \\ Konkoly Thege M. ut 15-17, H-1121 Budapest, Hungary \\ email: {\tt hareter@konkoly.hu}, {\tt paparo@konkoly.hu}}

\pubyear{2008}
\volume{301}  %% insert here IAU Symposium No.
\pagerange{119--126}
\jname{Precision Asteroseismology}
\editors{A.C. Editor, B.D. Editor \& C.E. Editor, eds.}

\begin{document}

\maketitle
\begin{abstract}The star HD 51844 was observed in CoRoT LRa02 as a Seismo target which turned out to be an SB2 system. 
The 117 days long light curve revealed $\delta$ Scuti pulsation in the range of 6 to 15\,\cd\ where four frequencies have amplitudes larger than 1.4 mmag and a rich frequency spectrum with amplitudes lower than 0.6 mmag. Additionally, the light curve exhibits a brightening event recurring every 33.5 days with a maximum of 3 mmag and a duration of about 5 days. Thus, this star can be considered as a heartbeat candidate. The radial velocities from spectroscopy confirmed an eccentric binary system with nearly identical masses and physical parameters. The brightening event of the light curve coincides with the maximum radial velocity separation showing that the brightening is in fact caused by tidal distortion and/or reflected light.
One component displays large line profile variations, while the other does not show significant variation. The frequency analysis revealed a quintuplet structure of the four highest-amplitude frequencies, which is due to the orbital motion of the pulsating star. 
\keywords{stars:individual:HD51844, stars:variables:$\delta$ Scuti, binaries:spectroscopic}
\end{abstract}

\firstsection

\section{Introduction}
Pulsating components of binary systems are of special interest because the masses of the components can be derived directly if the inclination angle $i$ can be constrained sufficiently. These masses can then be compared to theoretical predictions from evolution theory and pulsation theory. 

\cite[Thompson et al. (2012)]{Thompson12} propose a class of eccentric binary systems with dynamic tidal distortions found among {\it Kepler} light curves. They coined the term ``heartbeat stars`` for this class. The light variation allows to constrain orbital elements such as orbital period, inclination, argument of periastron and eccentricity.

\ds\ stars are located within the classical instability strip extended to the main sequence. They pulsate in p modes with typical frequencies ranging from 5 to 50\,\cd. For a detailed review we refer to \cite[Breger (2000)]{Breger00} and \cite[Pamyatnykh (2000)]{Pamyat00}.
  
\section{The CoRoT Light Curve}

\begin{figure}
   \centering
   \includegraphics[width=3.0in]{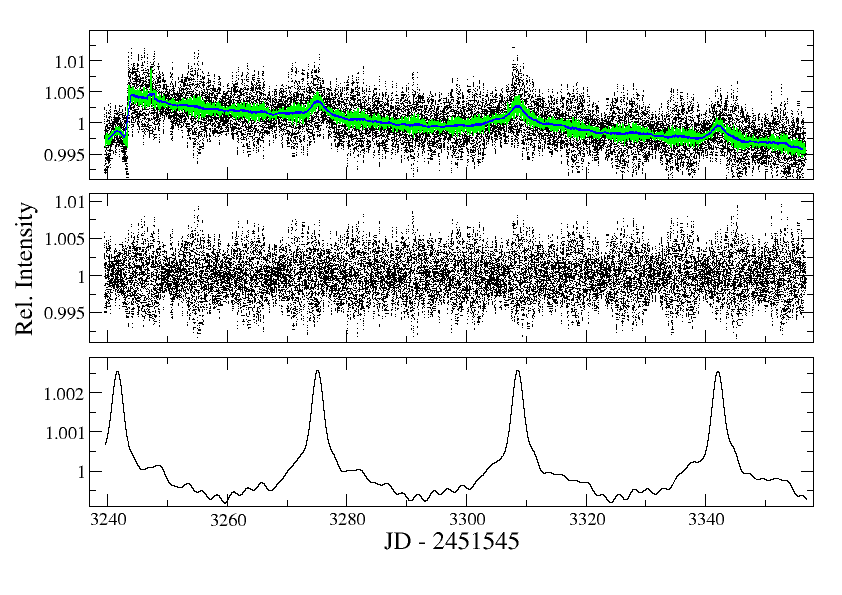}
      \caption{CoRoT N2 light curve (top panel, full dots) together with running averages (300 and 3000 points) to highlight the pulsation and the light variation due to the binary and instrumental effects. To obtain the pure pulsation light curve (middle panel), the binary model (shown in the bottom panel) was subtracted. Note that the y-axes are of different scales.}
         \label{fig-lc}
   \end{figure}

The binary HD\,51844 was observed in the CoRoT LRa02 for 117 days continuously. The high-quality light curve shows multi-mode \ds\ pulsation and a periodic brightening event.
Fig.\,\ref{fig-lc} shows the CoRoT light curve together with two running averages (300 and 3000 points) to illustrate the p-mode pulsation and the light variation due to the orbital motion of the stars. The outliers and jumps were removed in an iterative approach by involving prewhitening of the $\delta$ Sct type pulsation and correcting the residuals. The removal of outlying data points involved a 3-$\sigma$ clipping on the residuals rather than the original data. Since the pulsation frequencies are well separated from the binary light variation we constructed a simple binary light curve by calculating a fit from the first fifteen harmonics of the orbital frequency. 
  
\section{Pulsation and Frequency Modulation}

The subsequent frequency analysis was performed on the corrected data where the binary model was subtracted. In this paper we focus on the effects of the binary nature. A more detailed frequency analysis will be published elsewhere, though a brief summary shall be given below.

The frequency analysis using SigSpec (\cite{Reegen07}) resulted in more than 700 peaks using a significance limit of 6. The pulsation frequencies range from 5 to 15\,\cd\ where four frequencies have amplitudes larger than 1.4 mmag (in the integrated light from both components). These dominant frequencies (f$_1$ to f$_4$) are 12.213, 7.054, 6.943 and 8.141\,\cd , respectively. Around the two dominant modes (f$_1$ and f$_2$) there are closely spaced frequencies. f$_4$ is a single frequency outside of the groups and reveals an exact 3/2 ratio to the dominant mode. This ratio was also detected among a few frequencies with lower amplitude. 

The high-quality of the CoRoT Seismo data allows to detect frequency modulation due to the orbital motion. In an eccentric binary system a more complex structure of the side lobes is expected (\cite[Shibahashi \& Kurtz 2012]{sh12}). In our case we find around the four dominant frequencies first- and tentatively second-order side-lobes. The amplitudes of the latter range from 6.7 to 20 ppm. 
The average noise of the residuals after prewhitening is 1.8 ppm, which transforms to a S/N of 3.7 for the lowest-amplitude side-lobe. The amplitudes of the lower-frequency side-lobes ($a_{i-}$) are systematically higher than those of the higher-frequency sidepeaks.

\begin{table}
\caption{Frequencies, Amplitudes and phases used for the determination of theoretical RV curve.}             % title of Table
\label{tab-freq-split}      % is used to refer this table in the text
\centering                          % used for centering table
\begin{tabular}{lllll lllll lllll}        % centered columns (4 columns)
\hline                 % inserts double horizontal lines
$f_{c}$ & $f_{1+}$ & $f_{1-}$  & $f_{2+}$  & $f_{2-}$ &  $a_{c}$ & $a_{1+}$ & $a_{1-}$ & $a_{2+}$ & $a_{2-}$  & $\phi_{c}$  & $\phi_{1+}$ & $\phi_{1-}$ & $\phi_{2+}$ & $\phi_{2-}$ \\ 
\multicolumn{5}{c}{[\cd]}  & \multicolumn{5}{c}{[ppm]}  &  \multicolumn{5}{c}{[$-2\pi, 2\pi$]}\\% table heading 
\hline                        % inserts single horizontal line
12.213 & 12.243 & 12.183 & 12.274 & 12.150 & 2099 & 62 & 67 & 6.7 & 12.6 & 2.28 & -2.26 & 1.94 & -1.68 & 3.09 \\ % inserting body of the table
7.054  & 7.084 &  7.026 & 7.113 & 6.990 & 1953 & 35 & 48 & 9.7 & 8.1 & -3.10 & -1.81 & -0.41 &  -2.54 & 0.97 \\
6.943  & 6.972 &  6.912 & 7.007 & 6.988 & 1679 & 33 & 43  &   15 &  8.5 & -2.70 & -2.90 & 0.40 & 0.05 & 1.91\\
8.141 & 8.169 & 8.110 & 8.194 &  8.081 & 1371 & 25 & 50 & 13& 20 & -2.12 & -1.76 & -2.53 &  1.89 & 0.72 \\   
\hline                                   
\end{tabular}
\end{table}

\begin{figure}
   \centering
   \includegraphics[width=3.0in]{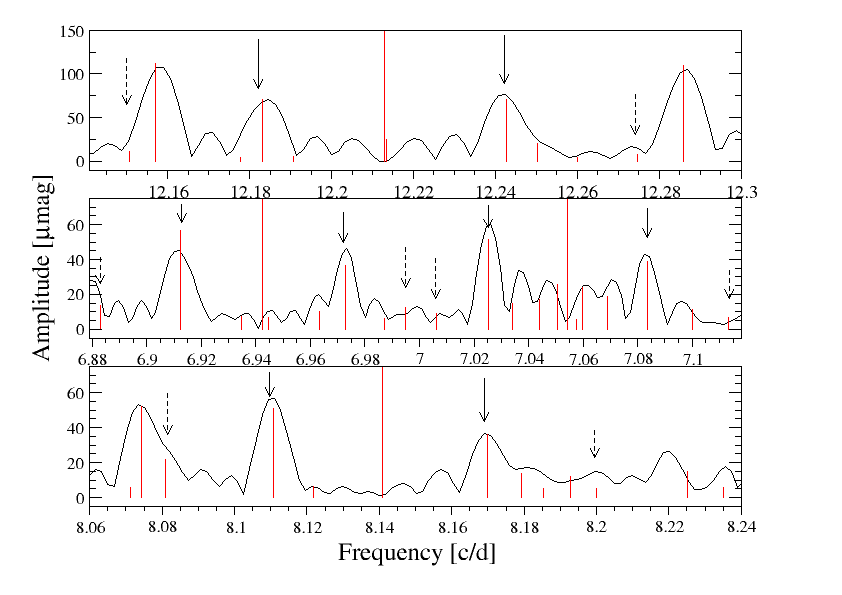}
      \caption{Identified candidate side-lobes at the four dominant frequencies. The full line shows the Fourier transform after prewhitening with the four dominant frequencies and the vertical full lines show the frequencies identified by SigSpec. The first-order side lobes spaced by 0.03\,\cd\ are marked by full-line arrows while the second-order side lobes are marked by dashed-line arrows.}
         \label{fig-freqmod}
   \end{figure}

Candidate frequencies for the second-order side peaks were found around the 4 dominant frequencies but 
no side-lobes were detected for the other frequencies.
The amplitudes of these second-order side peak candidates range from only 6.7 to 21 ppm. We show that for low-amplitude pulsators (highest amplitudes only $\approx$ 2 mmag) a noise level well below 1 ppm is required to recover the RV curve purely from photometry.

Fig.\,\ref{fig-freqmod} shows the structure of side-lobes around the 4 dominant peaks, where the corresponding modulated frequency was subtracted and Table\,\ref{tab-freq-split} lists the four dominant pulsation frequencies (column f$_c$), the identified side-lobes (f$_{i\pm}$), the corresponding amplitudes (a$_{i\pm}$) and phases ($\phi_{i\pm}$, zero point: beginning of CoRoT observations JD 2454784.565).

\section{The Orbit}

The binary nature of this star was discovered in the ESO LP185.D-0056 HARPS spectroscopy programme (led by E. Poretti). The light curve clearly shows a recurring brightening event with a period of 33.498 days. Thus, a spectroscopic campaign was organised with a good phase coverage that confirmed this period as the binary orbit period. These spectra include 19 HERMES and 5 McDonald spectra.

The radial velocities (RVs) were determined by fitting synthetic spectra to the observations using the software BinMag3 (by O. Kochukhov, www.astro.uu.ee/$\sim$oleg). First, the \vsini\ was determined by fitting synthetic spectra (both with \Teff\ = 6800\,K and \lg\ = 3.5) to the observations in which the lines of both components are well separated. In total 60 measurements were averaged and the mean values of \vsini\ = 41.4$\pm$ 1.5 and 41.7 $\pm$ 1.7\,\kms\, respectively, were determined. Thus, keeping the \vsini\ fixed for both components at 42\,\kms\ we determined the RV curves for all available spectra by using the same software. For each spectrum 16 windows were selected to fit the RVs. The standard deviations for the phases where the lines of the components are separated are 1.5\,\kms\ and for the phases where the lines are blended $\approx$ 3\,\kms.

\begin{figure}
   \centering
   \includegraphics[width=3.0in]{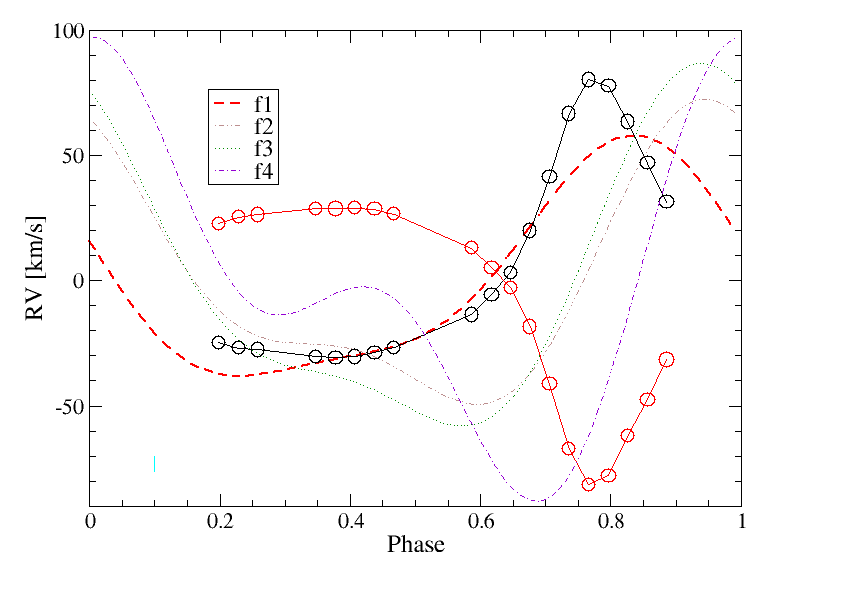}
      \caption{Comparison of the RV curves from spectroscopy to the RV curves from the frequency modulation. The spectroscopic RV measurements are corrected for V$_{0}$ and their errors are below 3\,kms, which is approximately of the order of the symbol size.}
         \label{fig-pincers}
   \end{figure}

The theoretical RVs curves calculated from the frequencies (Tab. \ref{tab-freq-split}) are compared to the observed RVs in Fig\,\ref{fig-pincers}. For f$_1$ the agreement is acceptable, while for f$_2$ to f$_4$ the agreement is poor. Because the second order side lobes have extremely low amplitudes they are prone to significant relative errors. Hence, we estimate that a noise level of less than 1 ppm in the Fourier domain is required to reconstruct the RV curve purely from photometry for such low-amplitude $\delta$ Sct stars. 

The brightening of the system is at maximum, when the RVs of the components are maximum. Thus, a tidal deformation and reflected light is likely to cause the brightening (``heartbeat effect'').

\section*{Acknowledgements}
\noindent 
M.H and M.P. acknowledge financial support of the ESA PECS project 4000103541/11/ NL/KML.
M.H is grateful to J. Benk\H o and L. Fosatti for valuable discussion.
We are grateful to the observers P. de Cat, P. Lampens, P. M. Arenal, J. Vos, and M. Rainer for collecting the spectra. 

%\bibliographystyle{../bibtex/aa}
%\bibliography{../hd51844-references.bib}

\end{document}